**Detection and optical imaging of induced convection under the action of static gradient magnetic field in a non-conducting diamagnetic fluid.**

Amit R. Morarka[a,b]
[a]Department of Physics, Abasaheb Garware College, Pune, INDIA.
[b]Department of Electronic Science, Savitribai Phule Pune University, Pune, INDIA.
Fax: +91-020-25699841; Tel: +91-020-25696060;
E-mail: amitm@electronics.unipune.ac.in, amitmorarka@gmail.com

†Appendix available: [Measurement of magnetic field and magnetic field gradients of the magnet and various configurations of magnets]

**Abstract:** The report elaborates for the first time visual observations of induced convections in a non-conducting diamagnetic fluid under the action of static gradient magnetic field in the absence of thermal gradients and the techniques employed to observe and record them. Suspension of Deionized (DI) and double distilled water and Lycopodium pollen grains was used as the fluid in a test tube. Permanent magnets having field strength of 0.12T each were used to provide the static gradient magnetic field. The suspension filled test tube was kept in a room temperature water bath. The convections were visually observed and recorded using travelling microscope attached with a web camera. Various geometrical configurations of magnets in the vicinity of test tube provided different types of magnetic gradient shapes. These gradients were responsible for the occurrence of different types of orientations in the convective flows in the test tube. Convections were observed over a range of fluid volumes from 0.2ml-10ml. The experimentally observed results provide proof of concept that irrespective of the weak interactions of diamagnetic fluids with magnetic fields, these effects can be easily observed and recorded with the use of low tech laboratory equipments.

**Introduction:** Effect of magnetic field on the dissolution kinetics of gas in distilled water was reported by [1]. The outcome of their study was that the oxygen gas dissolution rate was enhanced due to the magnetically induced convection in water. The onset of convection was attributed to two important factors viz; the presence of non-uniform magnetic susceptibility gradient formed by the permeation of oxygen gas phase into water phase and magnetic force given by equation (1). Their explanation was based on the assumption that the oxygen gas present at the surface of the sample under the pressure of $2 \times 10^4 Pa$, dissolves in the water sample. Thus a magnetic susceptibility gradient is created due to mixing of paramagnetic (oxygen) and diamagnetic (water) molecules. When the fluid was exposed to a gradient magnetic field, paramagnetic oxygen experiences the magnetic body force called as kelvin force [2, 3]. Mathematically magnetic force is given by equation (1). This body force is responsible for the onset of convection in their sample. Once the circulation is established, the circulating current under the surface of water enhances the dissolving rate of the molecular oxygen by carrying it upto the bottom of the sample container.

$$F = -\mu_o \rho \chi H (\partial H / \partial Z) \qquad (1)$$

Equation (1) explains the attractive and repulsive magnetic force on non-conducting fluids with positive and negative magnetic susceptibilities respectively. The magnetic force on a diamagnetic fluid is directed away from high magnetic field regions and is proportional to the square of the field gradient. The force acts in opposite sense in the case of paramagnetic fluids. [3] Shows that irrespective of the small magnitude of this force, it can be utilized to control thermal convection in pure water using high static magnetic field. Their immediate work [4] demonstrates that the force can be treated as the gravity analog (termed as: effective gravity) which produces magneto-thermal buoyancy in differentially heated fluids.

Based on the body force, discussions provided by [1] claims that the onset of the convections which are responsible for the enhancement of dissolution of oxygen is due to the susceptibility gradient created in the fluid by the presence of oxygen gas on the surface of the water.

Other report [5] showed through their numerical simulations that the gradient magnetic field either enhances or inhibits the thermal convections in the diamagnetic non conducting fluid (water). They concluded with two important results which stated (i) thermally driven convections are controlled by the application of the static gradient magnetic field having a critical value and (ii) above the critical value of the static gradient magnetic field; the thermally driven convections are completely replaced by the magnetically driven convections. Experiment such as [6] was performed in a sample volume of 0.3ml using non conducting diamagnetic fluid (water) along with a thermal gradient of 5°C. This system was exposed to gradient magnetic field having the magnitude of the product B(dB/dz) to be $1360T^2$/m. One of their results indicates the presence of small downward convection inside the sample under the heater. This is the state in the system in which the magnetic force dominates the buoyancy driven convections. Similarly [7] used the 3D computational modeling to obtained results which showed that the large magnetic forces can cause the onset of the axisymmetric magnetothermal convections.

In [5, 6 and 7], one common concept was considered that the volume magnetic susceptibility of the fluid is an implicit function of temperature, i:e; $\chi_v = \chi_g(t)$ where [($\chi_v = \chi_g*\rho$) and $\rho$ is the density of the fluid]. The described works by [5, 6 and 7] was based on the simultaneous interaction of the thermal and magnetic gradients in the fluid. None of them reported any observations on the onset of convection currents due to only static gradient magnetic field in the diamagnetic fluid (water).

Till date all the reported work related to the magnetothermal convection in diamagnetic fluids (water) was carried out using high field superconducting magnets, this is because water has a very small diamagnetic susceptibility value $-9.1 \times 10^{-5}$. For all practical purposes it was thought that the interaction of magnetic field with water will be too small to produce any cognizable effect on the macroscopic level using magnetic fields of the order of 0.5 T. Hence previous reports were unable to produce any experimental proofs about the direct observations of the occurrence of the convection in water under the application of static gradient magnetic field in the absence of the thermal gradients. The present report elaborates experimental findings about the onset of the convection which is independent of magnetic susceptibility and temperature gradients. A simple technique is employed to detect and capture magnetically induced convection in a diamagnetic fluid. The experiment was performed using DI water as the diamagnetic working fluid, rare earth magnets having 1.2kilogauss field strength[1] (Appendix), diode laser (<1mW), travelling microscope and a web camera to capture the convection flows. Absence of temperature gradient was measured using two calibrated RTDs.

---

[1] The field strength of the magnets was measured by placing the hall probe in physical contact with the flat surface of the magnet.

**Experimental:**

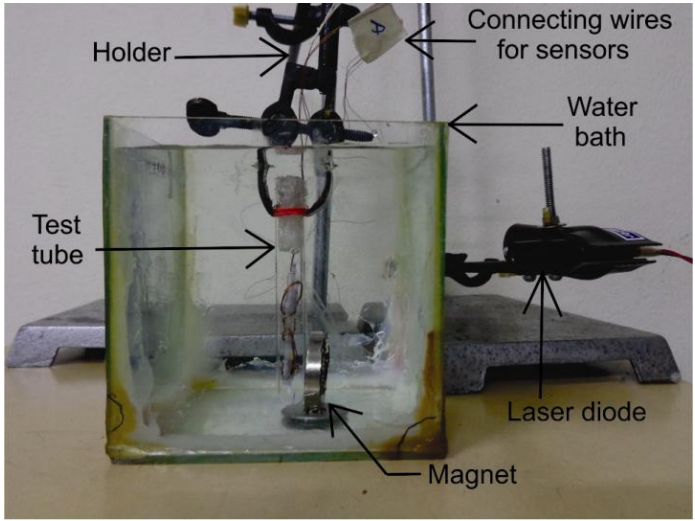

*Figure 1a: Image of the water bath setup used for the measurement of thermal gradients in the DI water in a test tube at room temperature. The same setup was used to show the onset of convection currents due to the interaction of static gradient magnetic field and the DI water.*

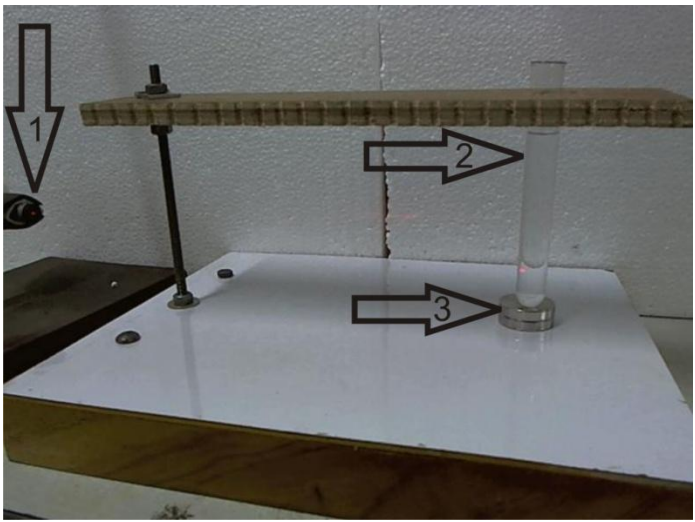

*Figure 1b. Image of the experimental setup used for the observation and recording of the induced convections due to applied static gradient magnetic field without the water bath. Arrows 1, 2 and 3 indicates the laser diode, test tube filled with the suspension and the two disk rare earth magnets kept below the test tube.*

The experiment was performed using distilled and DI water. For both the fluid types the results were identical. Hence for all further work, DI water was used.
A stock solution of lycopodium-water [8, 9] was prepared by adding 100μg of lycopodium powder in one liter of DI water. From that stock solution various volumes were taken in a glass test tube for the experiments. For every new experiment a fresh stock solution was prepared and used. A suspension from the stock solution was taken in the test tube. The test tube was left undisturbed for an hour so that if the water in the tube had any directional flows, they did die out.
The experimental setup is as shown in figure 1a and 1b. Figure 1a shows the experimental arrangement used for measuring the presence of temperature gradients and for observing magnetically induced convections in water at room temperature water bath. The bath had a water volume of 900ml. An airtight sealed 5ml test tube containing water and lycopodium suspension was kept inside the water bath. Two PT-100 RTDs were sealed inside the suspension without touching the inner walls of the test tube. The two RTDs were connected to the Agilent's six and half digit multimeter. They were alternately switched while taking noting down their resistance values. The accuracy of the multimeter was 0.001Ω for measuring the two RTD's resistance values. Similarly figure 1b shows a 10ml test tube filled with the suspension which was used to observe convections but without the water bath. The tube after filling was kept undisturbed for an hour at room temperature before starting the experiment. For observing the convection, a low cost and simple PIV technique which was based on the previously reported work [8, 9, and 10] was developed. The suspension was irradiated with a 630nm and having an output power less than one milliwatt laser diode. Motions of these pollen grains were visually observed and recorded using a 10X microscope with an attached web camera. The diode laser was fitted with a Plano convex lens of focal length 3cm, Figure-2 (Side and Top view). The use of lens facilitates variations in the laser beam diameter, which greatly simplifies the detection of the particles through the microscope or even with naked eyes. Though the output power of the laser is very low, it provided a high degree of illumination when observations were carried out in a room which either was completely dark or bright due to the absence or presence of ambient light.

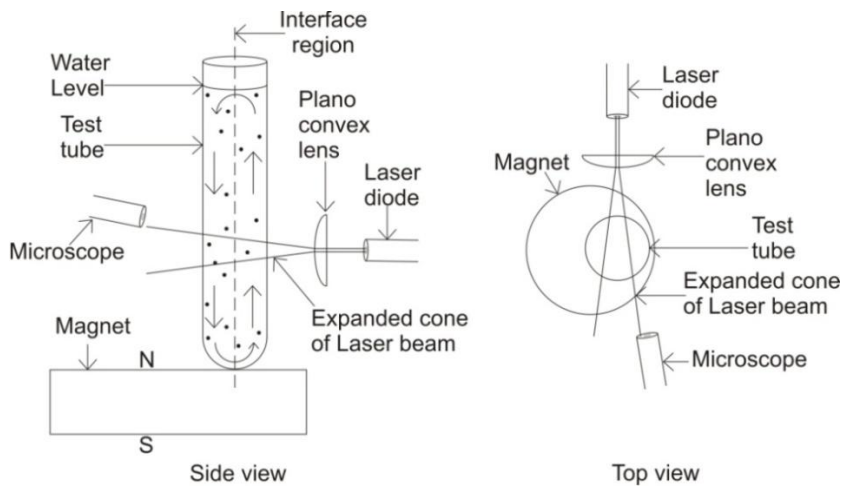

*Figure 2*
*Side View: Test tube filled with lycopodium water solution exposed to the magnetic field. Laser beam was diverged using a plano-convex lens. Convection currents are shown by the arrows. Black dots signify the suspended pollen grains.*
*Top view: The test tube is placed off center over the disk magnet. The Solid lines depict how the laser light traverses through the lens, test tube and into the microscope.*

In the side view of the figure 3, the test tube filled with the suspension was placed on the magnet. The convection flow starts as soon as the test tube is exposed to magnetic field.

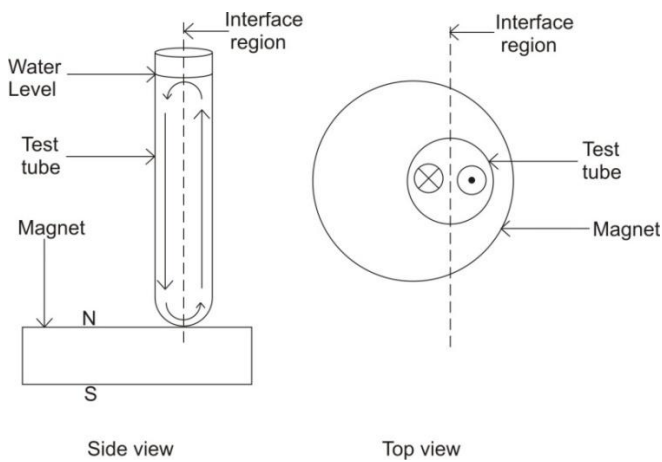

*Figure 3: Side View*
*Test tube is filled with the suspension is kept over the magnet. Dash line indicates the axis plane separating the upward and downward flow directions.*
*Top View*
*The cross and the dot in the circle indicate the convection flow direction going into the page and coming out of the page respectively with respect to the axis plane.*

Video clips of the convection flows were taken using the setup as described above. The obtained video clips were converted into jpeg images using Avidemux (freeware) software. The video clip was cut into numerous images each corresponding to one frame. Out of such a large collection of images, few were selected in such a way that when they were superimposed on each other to produce a single overlayed image. Thus the single image shows the evolution of the particle trajectory over certain period of time. Figures 4, 5 and 6 shows the time evolve images of upward flow, downward flow and interface region respectively.

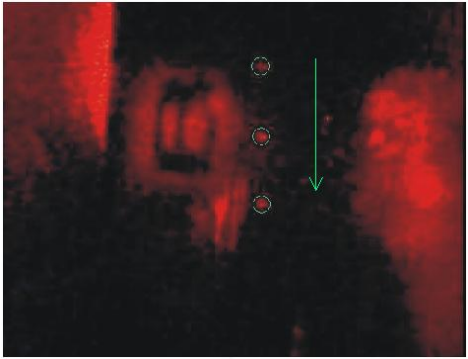

*Figure 4: Time evolved overlayed images of Upward flow in the convection. Green arrow indicates the trajectory of a single pollen grain which was caught in the flow.*

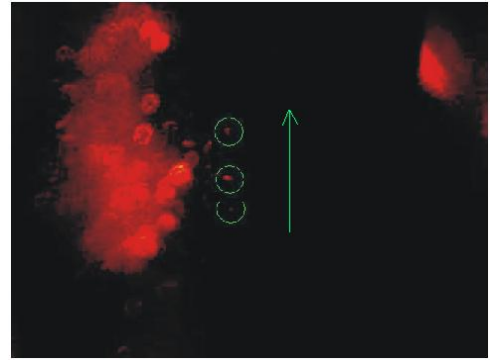

*Figure 5: Time evolved overlayed images of downward flow in the convection. Green arrow indicates the trajectory of a single pollen grain which was caught in the flow.*

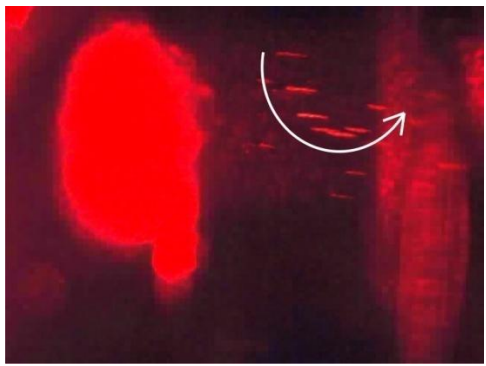

*Figure 6: Time evolved overlayed images of the trajectory of a single pollen grain traversing from upward flow into downward flow. White arrow indicates the path followed by the pollen grain. The flow velocity in this region is in the order of few micrometers per second.*

The effect of static magnetic field gradients on the formation of convection currents were studied by applying various magnetic field orientations with respect to the test tube. Figure 7 and 8 shows the dual and quad configurations of magnets respectively. In the dual configuration, the test tube filled with the suspension was kept in between the two disk magnets which were held in front of each other by two bolts. Similar to the previous technique, the microscope and the laser light were used to observe the convection. Observations were taken by directing the laser light and placing the microscope over the magnets, as shown in top view. The image depicts the dual configuration of magnets with the suspension filled test tube. Two disk magnets were used to enhance the field strength (Appendix Figure 3, 4) for the experiment.

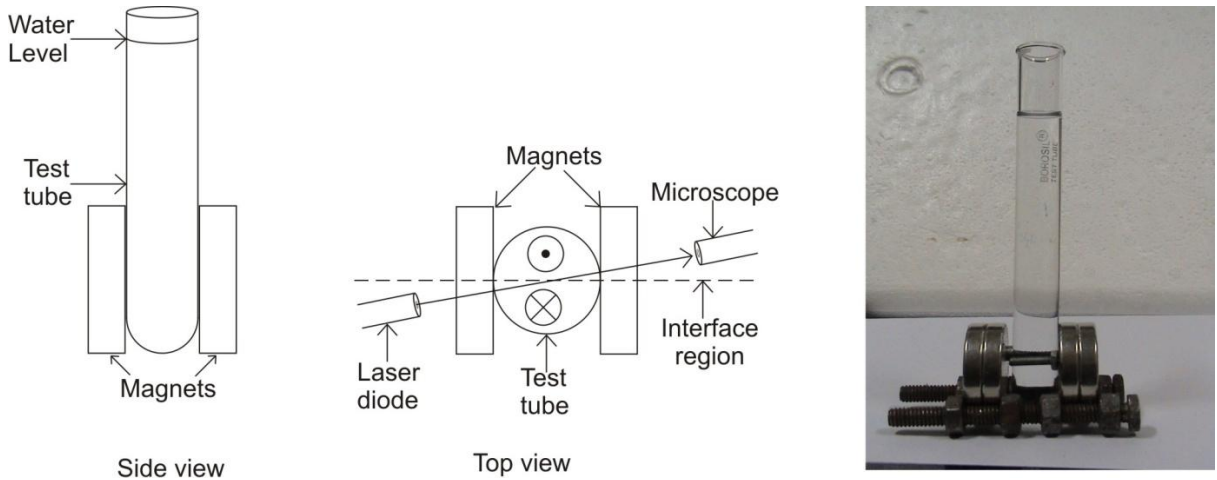

*Figure 7: Dual configuration of magnets*
*In side view, only unidirectional flow (Upward or Downward) is visible while observing the convection currents.*
*The top view depicts a setup which contains Laser beam and the microscope which was arranged in such a way that scattered laser light makes an angle with the interface (Dash line) plane making the interface plane visible. Cross and Dot in the circle represents the convection current direction going into the page and coming out of the page respectively.*
*The image on the right shows test tube filled with suspension positioned in a dual configuration of four permanent magnets placed and firmly tightened by screw and nuts.*

The quad configurations of magnets were made by placing four disk magnets as shown in Figure 8. Two types of configuration were used as seen in the top view I & II.

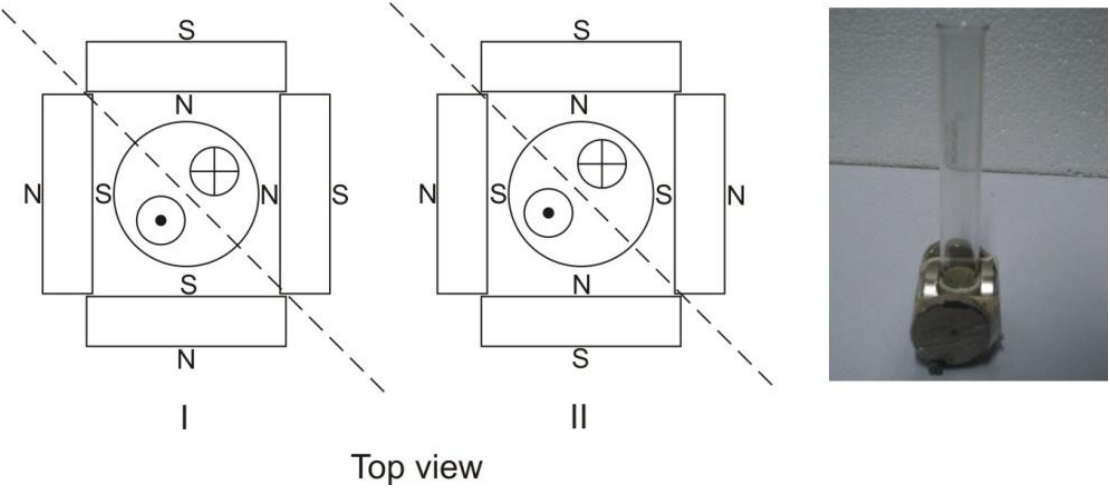

*Figure 8: Quad configuration of magnets*
*Four magnets are arranged in two types of configurations with respect to their orientation of poles with each other as shown in top view (I & II).*
*The convection flow directions are shown as explain before. As seen in configuration I & II, the flow directions are oriented in such a way that the axis plane (Interface region) is aligned with the diagonal of the rectangle formed by the placement of the four magnets.*
*The image on the right side shows arrangement of four disk magnets around the test tube.*

To understand the dependence of the convection on the physical positioning of magnet with respect to the test tube, the magnet was placed above 1cm from the horizontal surface and in the proximity of the test tube filled with the suspension as shown in Figure 9. Similarly another arrangement was made wherein the magnet was placed on the top opening of the test tube which was filled by the suspension upto its brim, as shown in Figure 10.

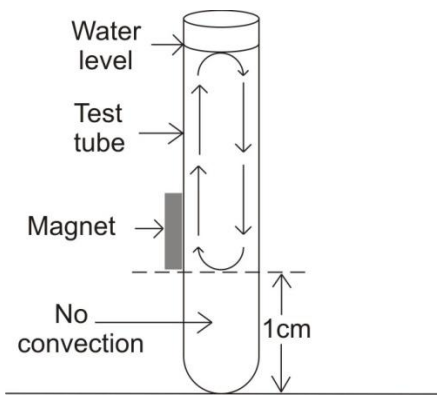

*Figure 9: Presence of convection currents only up to the geometrical dimension (Marked by Dash line) of the disk magnet.*

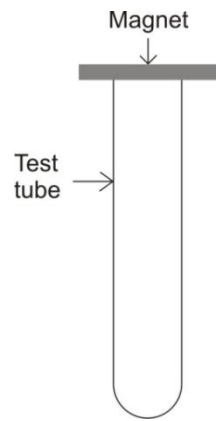

*Figure 10: Convection currents in the test tube do not occur when the suspension was exposed to the magnetic field from top of the test tube.*

It is a well-known fact that as the volume of the water in the test tube is reduced, temperature distribution over the entire water sample becomes isotropic. Hence to eliminate the cause of the convection currents due to presence of temperature gradient, 0.2ml of suspension was taken in the test tube, which was exposed to magnetic field as shown in Figure 11. Similarly to check whether the laser light is providing any heat energy to the suspension to initiate the convection, an arrangement shown in the Figure 12 was made. It consists of a test tube filled with the suspension with the disk magnet in its proximity but 1cm above the bottom side of the test tube. The laser was kept ON for one hour. The suspension was exposed to laser light and the magnetic field simultaneously.

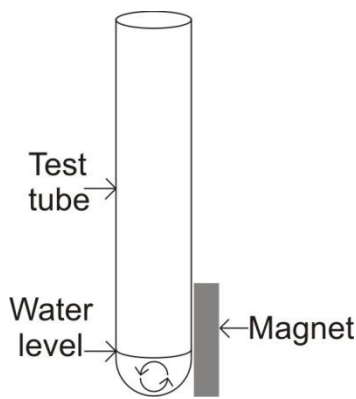

*Figure 11: Convection currents initiates in the 0.2ml suspension when exposed to the same static magnetic field. Arrow shows the direction of the convection currents.*

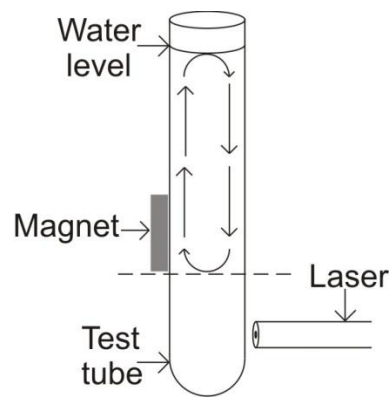

*Figure 12: Test tube filled with the suspension was exposed to laser radiation below the magnet for an hour. No convections were observed in the region were the laser light passed through the suspension.*

In Figure 13, the onset and flow directions were observed in the Thiele's tube. The tube was filled with suspension, left undisturbed for an hour and then the magnet was kept in the proximity of the arm of the tube as shown.

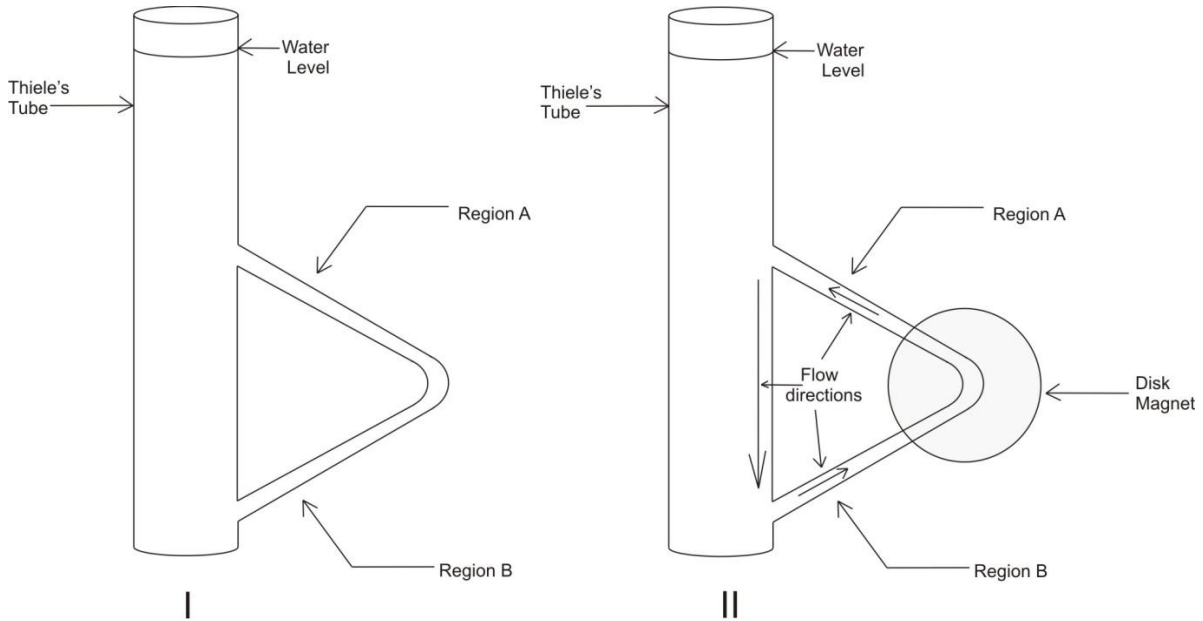

*Figure 13: Convection flow in Thiele's tube. Region A and B represents the two connected arms of the tube.*
*In the figure I, there are no convection flows in the absence of the magnetic field. As the disk magnet is placed at the joining of the two arms, figure II, the pollen grains are found to move along the direction indicated by the arrows, showing the convection flow in the tube.*

**Results and discussions:** Figure 1(a-b) shows sample tubes of 5-10ml kept in water bath and in open at room temperature. The convection currents were initiated as the suspension filled test tubes were exposed to gradient static magnetic field. The convection flow directions are depicted by the arrows inside the test tube in all the figures. The onset of convection current in the water is an instantaneous process when the suspension is exposed to static magnetic field which is in the vicinity of the test tube, as depicted in the figure 1(a-b). It was observed that the onset of the convection inside the test tube appears in the region which is closest to the surface (pole) of the magnet. Figure 2 shows how the placement of the disk magnet near the suspension filled test tube. It also shows the direction of the initiated convections, how it was observed and recorded. As shown in Figure 2 (Side and Top views), nearly 100% of the laser light goes through the test tube. Some amount is scattered by the wall of the test tube. Scattering through DI water is practically negligible. As the water contains few pollen grains suspended into it, scattered laser light due to these pollen grains was observed at a scattering angle over a (estimated) range of $5\text{-}10^0$ with respect to the unscattered diverged laser beam. The scattered light from pollen grains was collected through the microscope and in to the web camera. Due to the use of microscope the images obtained are inverted and are presented without any processing to their orientations. The camera was attached to a PC. Video recording freeware recorded all the videos of the convection currents. From the captured videos the maximum velocity of the convection currents were determined to be 2mm/sec. This velocity magnitude was deduced from a certain period of the entire convection flow video. In actual it was found that the flow velocity initially started slowly, then rising to its maximum velocity and then again slowly decaying to zero. The total period of the convection from its onset to complete decay depends upon the volume of water taken. The magnitude of the velocities in the convection increased as the applied field induction value was increased. This behavior is consistent with the force equation (1). The suspension which was initially in equilibrium due to hydrostatic pressure of the suspension column over it, acquires $(-\chi B^2/2\mu_o)$ amount of energy [11] in the magnetic field. This perturbation causes the suspension to start moving to minimize its energy. In doing so the suspension undergoes convective cycles. This convection dies out after a certain period of time which depends upon the volume of suspension, field*field gradient product and the viscosity of the fluid.

To find out whether the fluid sample under consideration contained any thermal gradients or not, an experiment was performed. The quantification of the experiment was done by measuring the difference in the resistance values of two similar calibrated PT-100 temperature sensors over 30-60minute of time interval. Since the resistance of any metallic conductor is an explicit function of its temperature, the resistance values of the two PT-100 were directly considered for measuring the existence of the thermal gradients in the suspension fluid.

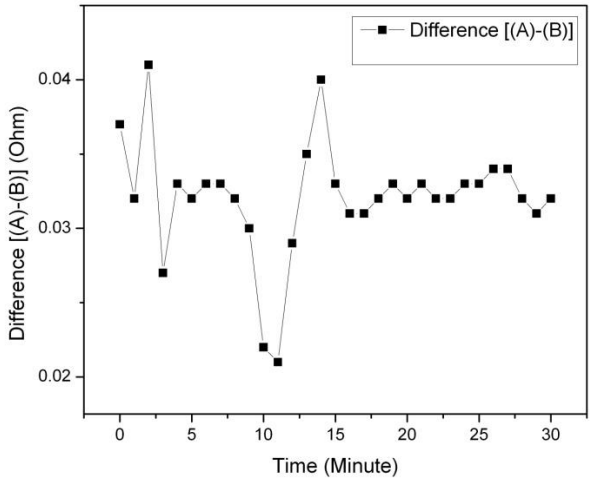

*Figure 14: Graph shows the difference in the resistance values of the two PT-100 sensors A and B. The calibration of these sensors was done by freezing them inside ice. Tap water of 200ml was used to make the ice.*

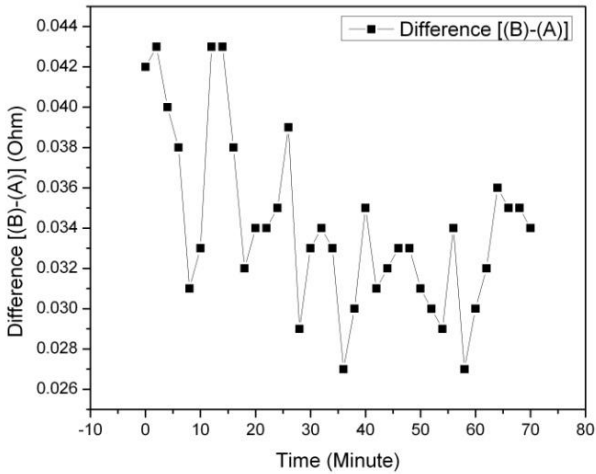

*Figure 15: Graph shows the measurement of difference between the resistances values of the two PT-100 sensors sealed inside the 5ml test tube containing suspension sample. The test tube was kept immersed inside a 900ml water bath which was at room temperature.*

Figure 14 shows the calibration of the two PT-100 sensors which were kept immersed inside the ice for 30minute. Ideally at the zero degrees Celsius (ice point) both the PT-100 sensors should show zero Ohm resistance hence, their difference should also be zero. But from the graph it can be seen that the average value of their differences oscillates around a value of 0.03Ω. The deviation from the zero resistance can be attributed to added resistance due to the connecting wires and the soldered points on the sensors. Similarly from the graph in the figure 15 it can be seen that the difference in the resistance values of the two PT-100 sensors kept immersed in the sample which is in a water bath at room temperature shows an average value of 0.03Ω for the difference in the two resistance values within the accuracy of 0.001Ω of the resistance measuring instrument. From the two graphs it can be concluded that there are no thermal gradients in the sample of the 5ml volume. When the same 5ml test tube in the water bath was exposed to the magnetic field from the disk magnet, the onset of convection was instantaneously seen as depicted in the figure 2.

As shown in Figure 3, when the test tube filled with the suspension was kept on the magnet at the same instant the convection flow begins in the direction as depicted in the figure. It can be seen in the Figure 3 side view that the test tube is positioned off centered on the disk magnet. This off centering provides maximum field gradient (Appendix Figure 3) to the suspension. It was easily observed that wherever the magnetic gradient was maximum the convection flow was initiated but in the upward direction. In this case the edge of the disk magnet has maximum field gradient value which causes the onset of the convection from the same edge and goes upward as shown in side and top view of the Figure 3.

The imaging of the convection flows was done using readily available and simple instruments. The trajectories of pollen grains captured in Figures 4, 5 and 6 are no different than any other regular convection flows observed inside a test tube which is heated from some arbitrary point. The only major difference is that the convection in this case arises due to the force experience by the water molecules in the space varying magnetic field. As the upward and downward flows pass each other at the center of the test tube, it creates a region in which the upward and downward flow mixes with each other. This region is called as the 'Interface region'. The mixing causes not only randomly oriented flows but also some directed flows. These directed flows were observed by looking at the trajectory of pollen grains. It can be seen in the Figure 6 that a pollen grain flowing in the upward direction with its velocity vector oriented slightly towards the interface region, crosses it and gets caught up into the downward flow.

Once the onset of the convections was verified, another fresh suspension was similarly exposed to magnetic field gradient but with a different magnetic field orientation. The test tube filled with suspension was positioned in between disk magnets as shown in Figure 7 and 8. The magnets were stacked in pairs of two on either side of the tube and were simply outfitted by using two screws and nuts. Appendix Figure 5-8 provides the image and magnitudes of the field gradient for this arrangement. In this configuration of magnets, there are two main field gradients. One field gradient is in the direction of the axis joining the faces of the magnets and the second is in the direction perpendicular to this axis. As the convection was observed, it showed a unique result as compare to the previously observed convection direction. The convection direction was perpendicular to the faces of the magnets as seen in the figure 7. In this case the interface region is perpendicular to the magnet faces. Since the field*field gradient product (Appendix Figure 9-10) is maximum in the direction perpendicular to the interface line, the convections were induced as depicted in Figure 7.

Similarly another arrangement of magnets which contained four magnets was made. As shown in Figure 8, magnets were arranged in such a way that they covered the test tube from all the sides (Appendix Figure 11). In this case the interface line coincides with one of the geometrical diagonal of the square that was formed by the placement of the four magnets.

In all of these observations one thing was common that the onset and the behavior of the convection were independent of the polarity of the poles of the magnet. Following on the similar lines, magnet was placed at various positions near the test tube. Figure 9 and 10 shows another two setups. As the formation of convection is cause due to the competing forces from magnetism and fluid hydrostatic pressure, the convection flows occurs only upto the region where the physical diameter of the magnet extents. The entire column of the fluid was unperturbed below the dashed line, as shown in the Figure 9. The convections covered the entire length of the column above the dashed line upto the maximum water level in the test tube. In the present phenomenon of formation of convection in a static gradient magnetic field, if the magnetic force is kept parallel and in the same direction to that of the gravitational force pointing towards the earth, convective flow does not occur. Figure 10 shows an arrangement where the magnet is kept on the open end of the test tube. The test tube was filled completely upto it's opening where the fluid was in physical contact with the magnet surface. The system was observed continuously before and after the magnet was placed over it. The fluid did not show any formation of convective flows irrespective of the time for which the magnet was placed over it.

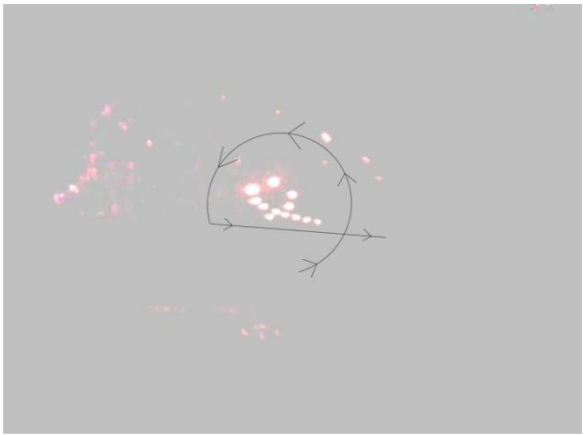

*Figure 16: Time evolved overlayed images of a single pollen grain caught up in the convective flow. The black line with the arrows on it indicates the trajectory of the particle.*

In a fluid system having volume of 10ml, temperature gradient driven convections can easily be misinterpreted as the magnetically induced convections. To check whether the convections in the water are truly arising due to the effect of static gradient magnetic field, 0.2ml of suspension was taken in a test tube. As shown in figure 11, initially the system was recorded without placing the magnet. The drift velocity in the system was observed and recorded for 10 minute. At the end of 10 minute without disturbing the test tube, magnet was placed under it. The video was captured for 40 minute. The observations and recording was started in the absence of magnetic field, it was observed that the pollen grains were moving very slowly in random directions. After 10 minute a magnet was kept near the test tube. A very slow but gradual change in the position of the randomly moving pollen grains was observed and recorded. Over the period of time the gradual change in the position of the pollen grain traced a circular path and then it vanished from the view, indicating the presence of convective flow in the system. The captured video was then split into single frame JPEG images using Avidemux (Version 2.5) software. Out of all the obtained images, 66 images (30 frame interval) were selected. Using these images a time evolution montage of the motion of the pollen grain was created using the ImageJ software. Figure 16 shows the time evolve image of the pollen grain over the entire period for which the suspension was under the influence of the magnetic field.

Another experiment was down to find out whether the laser light imparts thermal energy into the suspension or not. Figure 12 shows the experimental arrangement. The convection flows occurred from the dashed line and into the entire above column of the suspension. There were no signs of any kind of ordered motion of the pollen grains in the column below the dashed line where the laser light source was kept 'ON' for an hour. The simultaneous existence of convective flows and stable fluid with no flows above and below the dashed line shows that the laser light source does not provide any cognizable thermal energy in to the system.

Based on the obtained results of how the convection behaved in the various arrangements of the magnets, a question was raised; can water be circulated through a close loop system? An attempt to answer this question was made by using a Thiele's tube. A Thiele's tube is made-up of glass having a handle (arm) as shown in the figure 13 (I). This tube was used because of its typical geometry. The arm has a volume of approximately 4ml and the remaining part of the tube (resembles test tube) has a volume of 61ml. As the tube was made for utilizing it as a rapid and uniform temperature bath through the use of convective flow of the fluid through the arm and into the big tube, it served the purpose in the proposed experiment.

As shown in the figure 13 (II), the tube was filled with the suspension. The tube was left untouched for one hour to stabilize the system. After an hour, scattered laser light from the regions (A) & (B) were observed using a microscope to verify that there are no directed flows acquired by the suspension. Once that was verified, a magnet was placed exactly near the corner of the arm of the tube. As speculated, the flow of suspension was observed over the same regions (A) & (B). It was found that water was flowing in from the opening at the bottom of the arm then going through the regions (A) & (B) and then finally enters into the bigger tube through the opening at the top. As shown in the figure 13 (II), the continuous flow in and out of the small arm and into the big tube showed the convective motion acquired by the fluid in the Thiele's tube. Similar to the previous observations, the fluid flow ceases after some period of time even though the magnet is not removed.

As per the report of [1], the enhancement of rate of dissolution of oxygen was attributed to the onset of magnetically induced convections in the water due to susceptibility gradient created in the fluid by the presence of pure oxygen atmosphere at the surface of the fluid. They calculated the convective flow velocity using the assumption that there exists a magnetic susceptibility gradient in the water. As per their experimental conditions, the concentration of oxygen at the surface of the water is saturated. It decreases gradually towards the bottom. This creates the magnetic susceptibility gradient as the oxygen is paramagnetic whereas the water molecule is diamagnetic. They used an electro-chemical oxygen sensor to measure the DO contain of the sample. Their reported measurements do not reflect that a thorough measurement was done at various levels in the fluid to claim the existence of the susceptibility gradient. A simple comparison of the volume of the sample and the size of the probe would provide a roughly good estimate of how under any normal experimental conditions, the probe will be unable to register any concentration gradients in the sample. The measured value was an average value of the concentration of the oxygen dissolve in the sample. Hence, it is incorrect to assume that there is any cognizable concentration gradient of oxygen gas in the sample and hence, the magnetic susceptibility gradient will also be negligibly small which will not directly contribute to the onset of the convections.

To support this discussion, an experiment was carried out by degassing [12] the suspension using nitrogen ($N_2$), a diamagnetic molecule, at 1 atmospheric pressure having flow rate of 25ml/sec for two hour. $N_2$ molecule is three fold less diamagnetic than water molecule. After two hour, the test tube was kept undisturbed for an hour under the pressure of one atmosphere of $N_2$ gas. As the magnetic field was applied as discussed earlier, the onset of convection currents were observed exactly similar to the convections observed in the fluid which was kept in open air. The velocity of the convection flow was measured in both the cases which came out to be of exactly same value as described in initial experiments. Hence, onset of convections cannot be attributed to the concentration gradient and magnetic susceptibility gradient respectively.

In all the cases the experiment was carried out at room temperature and repeated at various times over two years. As the volume of the water was also very small, it is believed that thermal gradients, if any, are too small to setup convection current. This is verified by the absence of ordered motion in the water column in the absence of magnetic fields. The occurrence of convections is still unclear but it can be attributed to Lenz's law for induced magnetism. The water molecule is diamagnetic in nature which implies that the water molecule has no net magnetic moment associated with it. Any external applied magnetic field will induce a magnetic field in the molecule (Lenz's law) but which will oppose the applied field. Thus under the action of gradient force, the resultant effect will move the diamagnetic molecule away from the high field region to the lower field region. As this motion of water molecules is additive which further builds up to macroscopic instability in the fluid. To nullify this instability the fluid undergoes rearrangement in the position of the molecules with respect to the external applied field. In doing so a convectional flow is setup in the fluid. As the system attains new state of equilibrium, the convections die out over a period of time depending upon the volume of the fluid. Onset of convections, average maximum flow velocity and quenching are directly dependent on viscosity of the diamagnetic fluid. All the observations were exactly same in the case where the sample fluid was double distilled water.

Based on the above discussions, the increase rate of oxygen dissolution in the water reported by [1] can now be said to be an effect which is caused by the onset of the convection currents which are induced due to the magnetic force acting on the water molecules when they are exposed to static non uniform magnetic fields.

**Conclusion:** It was demonstrated that static magnetic gradient field can induce convection in DI and double distilled water. The phenomenon was observed using simple apparatus. Dual and quad arrangements of the magnets showed completely new dynamic orientations of the convection in the test-tube. Assumption of obtaining convectional flow in the closed loop system using Thiele's Tube was verified by applying an identical magnetic field. Moreover the convections were established directly by virtue of the interaction of water molecules with the magnetic field, as the experiment is carried out in ambient isothermal conditions. The convections are generated isothermally was supported by the findings of convections in 0.2 ml of water at room temperature and the absence of thermal gradients in the sample volume. Heating effect of the laser light source does not contribute to the convections in the fluid. Onset of convections was also observed when the suspension was degassed using $N_2$ gas. A model has been discussed for the formation and dissipation of the convection currents in the non-conducting diamagnetic fluid suspension. Further investigations are underway to quantify and model the observe effects. Applications of the observed phenomenon are also being explored.

The presented experiment is based on the simple techniques. The results thus obtained prove the feasibility of observing such weak interactions between diamagnetic fluid and magnetic fields successfully.

**Acknowledgement**
The author wish to thank Dr. Subramaniam Ananthakrishnan, Mr. Golum shaifullah, Department of Electronics Science and Dr. Bhalchandra Pujari, Center for Modeling and Simulations, Savitribai Phule Pune University for providing laboratory facility, their valuable time and advice for the discussions on the topic from time to time.

# Appendix

As shown in Figure 1 and 2, two disked shaped (diameter – 30mm, width – 5mm) permanent magnets are stacked together to get 0.25T magnetic field strength over the surface of the magnet. Figure 3 shows the magnetic induction values measured from the surface of the magnet upto a distance of 30 mm away, indicated by the arrow pointing away from the surface of the magnet. The slope value was calculated for the two points X and Y in the graph. As the diameter of the test tube used was 10mm, the distance taken for the calculation of the gradient between the two points was just over 10mm.

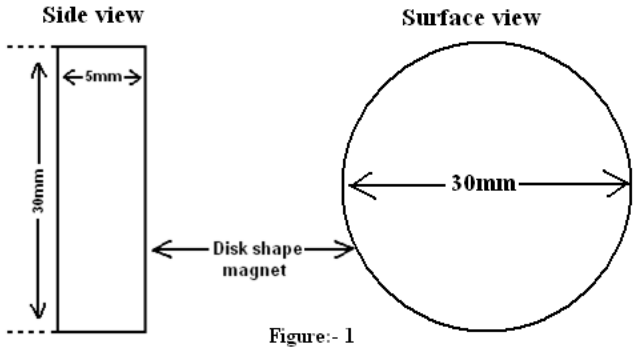

*Figure 1: The schematic showing dimensions of the disk shape permanent magnet.*

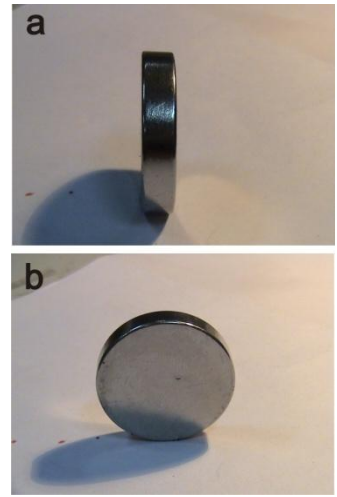

*Figure 2: (a) Side view & (b) Front view of the permanent magnet.*

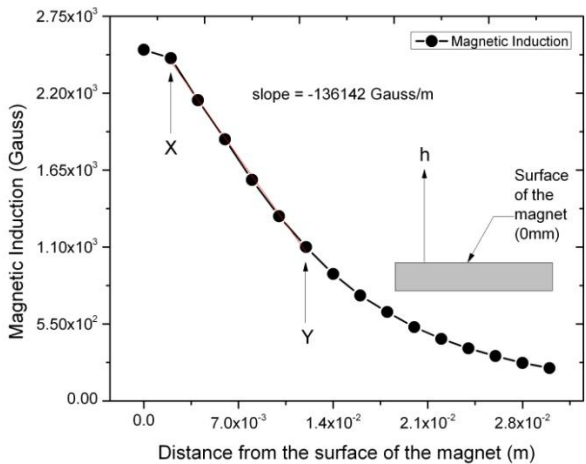

*Figure 3: Magnetic induction values at different points from the surface upto the distance of 30 mm. This length is designated by 'h' on the graph.*
*Slope = -136142 Gauss/m.*

Figure 4 shows how the magnetic induction varies on a surface of a disk magnet from edge to edge. The nature of the graph shows how steep are the magnetic gradients near the magnet surface. Hence, the effect of induce convection currents in the water due to magnetic force Fmag = $\mu_0 \chi \rho H (\partial H/\partial z)$, can be seen with the disk magnet due to its significantly large product value of field and field gradient. The effect was observed prominently when the tube was kept near the edge of the magnet, implying the dependence of the large magnetic gradient value.

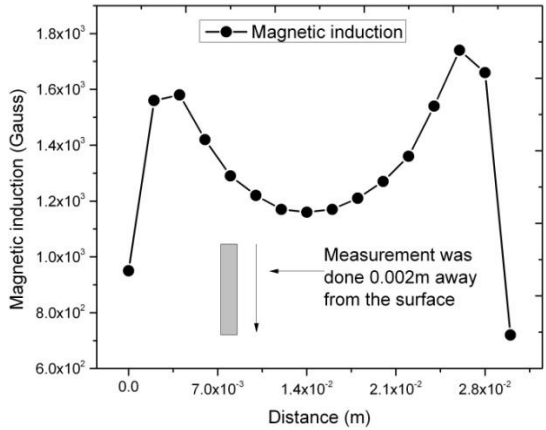

*Figure 4: Magnetic field strength of a disk magnet at a distance of 2mm from its surface as measured by gauss meter. The measurement was done across its diameter from one edge to anther edge.*

The dipole configuration consists of two magnets placed 30mm apart in attraction mode as shown in Figure 5. Two magnets were stacked together on either side. As the measurement probe was moved across from one surface of the magnet on to the surface of another magnet following the line of measurement, we obtain a minimum in the magnetic strength of the system which can be seen in the Figure 6.

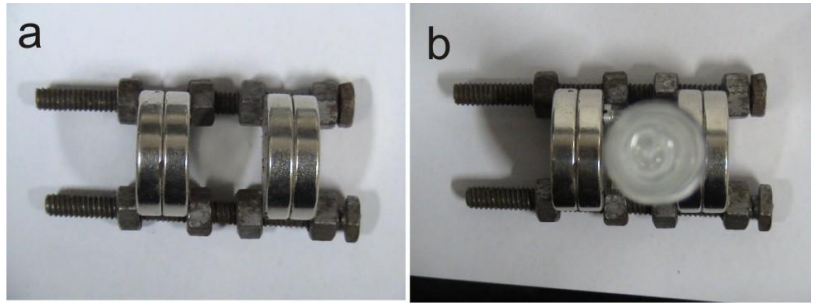

*Figure 5: It shows an image of the top view of the dual configuration of magnets. In the image 'a & b', the dual configuration of magnets with and without test tube are shown. Two magnets on each side are stacked together to enhance the magnetic field strength in their gap.*

Another set of magnetic field strength measurement was done in the same configuration. The measurements were taken along a line which is at a half the distance between the two magnets. Figure 7 shows the magnetic induction measurement done over the two arrow headed line indicated in the graph. On this line the magnetic induction has a maximum value at the center of the configuration and symmetrically decreases towards either sides of this maximum value.

*Figure 6: In the graph a small schematic shows the direction (Two headed arrow) along which the Magnetic induction was measured. Separation between the magnets was 18mm. N & S indicates the respective poles of the magnet. Two magnets were stacked together on either side to increase the field strength in the region.*

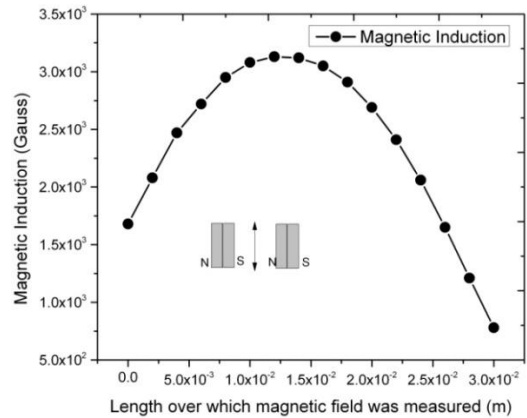

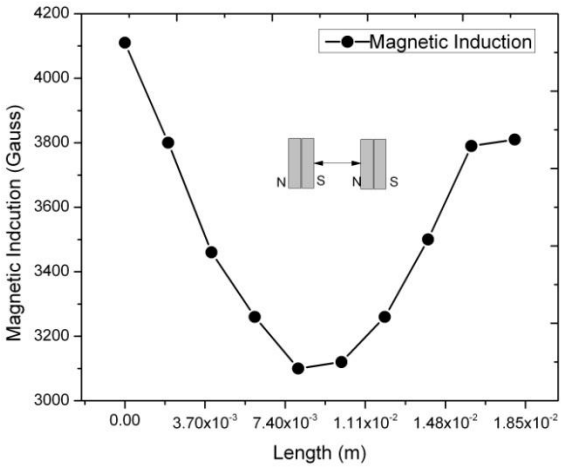

*Figure 7: Magnetic induction was measured over the two arrow headed line indicated in the graph in between magnets kept 18mm apart. N & S indicates the respective poles of the magnet.*
*Two magnets were stacked together on either side to increase the field strength in the region.*

Figure 8 show the graph of field gradient along the two headed arrow line. The field gradient was calculated from the graph of Figure 7. The gradient is nearly zero at the center and maximum near the two arrows.

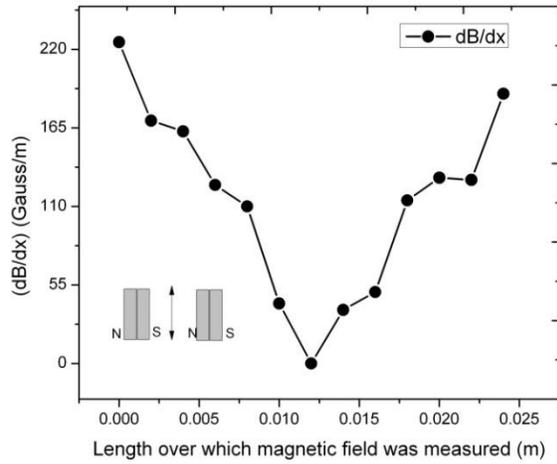

*Figure 8: Graph of field gradient versus length. It was measured over the line having two headed arrow. The two headed arrow line is at a half the distance between the two magnets.*

Figure 9 shows the graph of product of field and field-gradient {B*(dB/dx)} versus the distance over which it was measured. Comparing graphs from Figure 7 and 9, the minima of the product lie at the central point in between the magnets. The product value increases rapidly and symmetrically towards either side of the minima. This also indicates that the diamagnetic force experience by the water molecules will be maximum at the points where the product value is maximum. Hence, the onset of convections was observed when the suspension in a test tube was exposed to a magnetic configuration having a field field-gradient product distribution as shown in the Figure 9.

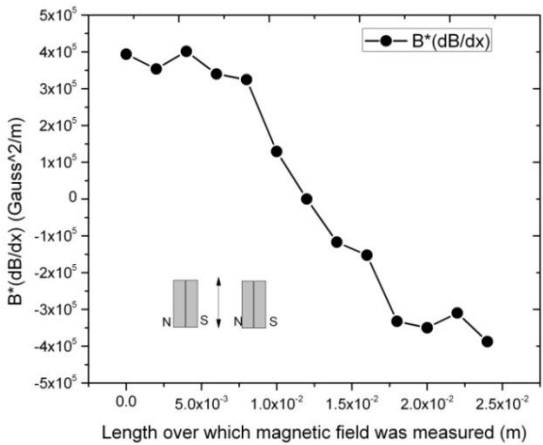

*Figure 9: Graph of product field-field gradient versus length. It was measured over the line having two headed arrow. The two headed arrow line is at a half the distance between the two magnets.*

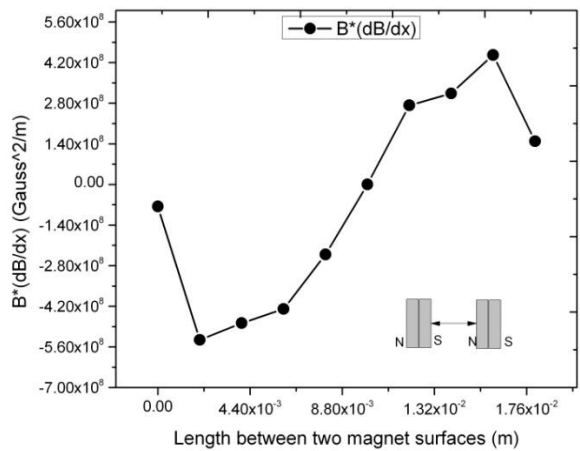

*Figure 10: Graph of product of filed field gradient versus length. It was measured over the line having two headed arrow. The two headed arrow line the distance between the two magnets.*

The quad configuration of magnets was made as depicted in Figure 11. Four magnets were attached to a test tube using single and double sided self adhesive tapes. The central cylindrical cavity formed due to the tapes thus had diameter little bigger than diameter of the test tube. This facilitated ease of placing of the test tube filed with suspension inside the quad configuration of magnets.

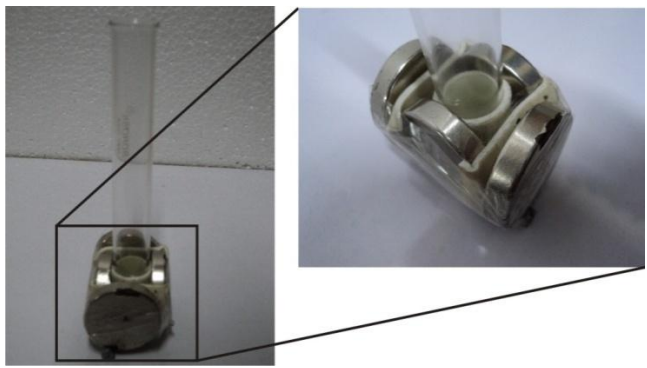

*Figure 11: Four magnets were placed together by using self adhesive double and single sided tapes. The inset shows the close-up view of the system.*